\documentclass[fleqn,usenatbib]{mnras}

\usepackage{newtxtext}
\usepackage{txfonts}

\usepackage[T1]{fontenc}

\newcommand{\beq}{\begin{equation}}
\newcommand{\eeq}{\end{equation}}

\newcommand{\kB}{\mbox{$k_\mathrm{B}$}}

\newcommand{\gcc}{\mbox{g~cm$^{-3}$}}
\newcommand{\msun}{\mbox{M$_{\odot}$}}

\usepackage[dvipsnames]{xcolor}

\usepackage{graphicx}	

\title[Thermal evolution of soft X-ray transients]{Thermal evolution of
neutron stars in soft X-ray transients with thermodynamically consistent
models of the accreted crust}

\author[A. Y. Potekhin,
M. E. Gusakov,
A. I. Chugunov]{
A. Y. Potekhin,\thanks{E-mail: palex@astro.ioffe.ru}
M. E. Gusakov,
A. I. Chugunov
\\
Ioffe Institute, Politekhnicheskaya 26, St. Petersburg, 194021, Russia 
}

\date{Accepted XXX. Received YYY; in original form ZZZ}

\pubyear{2023}

\begin{document}
\label{firstpage}
\pagerange{\pageref{firstpage}--\pageref{lastpage}}
\maketitle

\begin{abstract}

Thermal emission of neutron stars in soft X-ray transients (SXTs) in a
quiescent state is believed to be powered by the heat deposited in the
stellar crust due to nuclear reactions during accretion (deep crustal
heating paradigm). Confronting observations of SXTs with simulations
helps to verify theoretical models of the dense matter in the neutron
stars. Usually, such simulations were carried out assuming that the
free  neutrons and nuclei in the inner crust move together. A recently
proposed thermodynamically consistent approach allows for independent
motion of the free neutrons.  We simulate the thermal evolution of the
SXTs within the thermodynamically consistent approach and compare the
results with the traditional approach and with observations. For the
latter, we consider a collection of quasi-equilibrium thermal
luminosities of the SXTs in quiescence and the observed neutron star
crust cooling in SXT MXB 1659$-$29. We test different models of the
equation of state and baryon
superfluidity and take into account additional heat
sources in the shallow layers of neutron-star crust (the shallow
heating). We find that the observed quasi-stationary thermal
luminosities of the SXTs can be equally well fitted using the
traditional and thermodynamically consistent models, provided that the
shallow heat diffusion into the core is taken into account. The observed
crust cooling in MXB 1659$-$29 can also
be  fitted in the frames of both models, but the choice of the model
affects the derived parameters responsible for
the thermal conductivity in the crust and for the shallow heating.

\end{abstract}


\section{Introduction}
\label{sect:intro}

Many neutron stars reside in binary systems with a lower-mass companion
star (low-mass X-ray binaries, LMXBs) and accrete matter from the
companion. Some of the LMXBs, called soft X-ray transients (SXTs),
accrete intermittently, so that accretion episodes (outbursts) alternate
with periods of quiescence. During an outburst, the emission is
dominated by the accretion disk or a boundary layer (e.g.,
\citealt{InogamovSunyaev10,GilfanovSunyaev14} and references therein).
In quiescence, the accretion is switched off or strongly suppressed and
the luminosity decreases by several orders of magnitude (see, e.g.,
\citealt*{WijnandsDP17} for a review).

The accretion during an outburst leads to nuclear reactions in the crust
accompanied by heat release. In particular, the heat is produced as the
crust matter is pushed inside under the weight of newly accreted
material, which is known as the deep crustal heating scenario
\citep{Sato79,HaenselZdunik90}. Once an SXT turns to
quiescence, the accumulated heat leaks through the surface. The so-called
quasi-persistent SXTs have long outbursts (lasting months to years) that
are sufficient to appreciably warm up their crust. During periods of
quiescence between the outbursts, the thermal relaxation of the
overheated crust can be observed through X-ray emission from the surface
(e.g., \citealt{WijnandsDP17} and references therein). 

An analysis of observations of the post-outburst cooling allows one to
constrain the thermal conductivity and heat capacity of the crust (e.g.,
\citealt{Rutledge_02,Shternin_07,PageReddy13}). 
However, such an
analysis can be complicated.
Namely, the light curves of some SXTs in quiescence
can be reproduced within the deep crustal heating scenario
but require the so-called `shallow heating' by some additional energy sources at relatively 
low 
densities
\citep{BrownCumming09}.
Other SXTs may
require some residual non-monotonic accretion during quiescence
(e.g., \citealt*{TurlioneAP15}).

A traditional approach to  calculation of the equation of state (EoS)
and the heat release in the crust
\citep{HaenselZdunik90,HaenselZdunik03,HaenselZdunik08,Fantina_18} is
based on the assumption that free neutrons move together with the nuclei
during accretion.  Recently, the theoretical models of nuclear
transformations during accretion, accreted crust composition and deep
crustal heating have been revised by
\citet{GusakovChugunov20,GusakovChugunov21} and
\citet*{Shchechilin_GC21,Shchechilin_GC22},  who found that free
neutrons  may  redistribute independently of nuclei throughout the inner
crust and core  so as to satisfy the  hydrostatic and diffusion
equilibrium condition, $\mu_n^\infty={\rm constant}$, where
$\mu_n^\infty$ is the chemical potential of free neutrons,  redshifted
to the reference frame of a distant observer. In the major part of the inner crust,
where neutrons are superfluid, this condition is necessary  for
hydrostatic equilibrium of the system. The corresponding equilibration
timescale  is very short (of the order of the hydrodynamic timescale). 
In the rest of the inner crust, where neutrons are normal (in
particular, in the small region  near the outer-inner crust interface),
this condition is necessary for the diffusion equilibrium of free
neutrons. The associated timescale of reaching the equilibrium  is also
small (much smaller than the accretion timescale;
\citealt{GusakovChugunov20}).

As a result of implementing the
hydrostatic/diffusion equilibrium condition,
the EoS of the inner crust turns
out to be closer to the EoS of the ground-state (`catalysed') matter,
while the heat release inside the inner crust appears to be smaller than
predicted by the traditional models.
In this paper we examine the impact
of these results on the thermal evolution of the SXTs and discuss some
consequences for the analysis of the SXT observations.

The basic characteristics of the employed accreted crust models are
outlined in Section \ref{sect:accrust}. Section~\ref{sect:long} is
devoted to the modelling of the equilibrium thermal luminosities of the
SXTs as functions of the mean accretion rates in the frames of the
traditional and thermodynamically consistent models. We
also pay attention to the effects of the shallow heating, the neutron
star EoS and baryon superfluidity on these luminosities. In
Section~\ref{sect:MXB}, we apply the different crust models to
simulate
the thermal evolution of a neutron star in an SXT during
and after an outburst and compare the calculated light curves with
observations, using SXT MXB 1659$-$29 as an example.  Our conclusions
are summarized in Section~\ref{sect:concl}.

\section{Accreted crust models}
\label{sect:accrust}

In the previous papers \citep*{PotekhinCC19,PotekhinChabrier21}, we
studied the thermal evolution of the SXTs for different EoSs,
neutron-star masses $M$, nucleon superfluidity models and the total
accretion time $t_\mathrm{acc}$, employing the theoretical models of the
accreted crust \citep{HaenselZdunik08,Fantina_18}, which followed the
traditional paradigm, in which free (unbound) neutrons and nuclei move
together in the inner crust. The results obtained with these two models
proved to be similar, so in the present study we will limit ourselves to
considering the more recent one \citep[][hereafter F+18]{Fantina_18} as
a representative of the traditional model.

The thermodynamically consistent accreted crust models, constructed by
\citet[hereafter GC]{GusakovChugunov20,GusakovChugunov21}, depend on the
pressure $P_\mathrm{oi}$ at the interface between the outer and inner
crust, where free neutrons appear. Because of 
their
diffusion, $P_\mathrm{oi}$ may differ from the pressure at which neutrons
start to drip out of nuclei. The value of $P_\mathrm{oi}$ is currently unknown,
but the thermodynamic consistency and other arguments suggest that it
should be
between the minimum $P_\mathrm{oi}^\mathrm{(0)}$, which corresponds to
the case where no heat is released in the inner crust (i.e., in the
region with
$P>P_\mathrm{oi}$), and maximum $P_\mathrm{oi}^\mathrm{(cat)}$, which
roughly equals the neutron-drip pressure in cold catalysed matter.
As in
the traditional models, the crustal
heating  
occurs at certain pressures (or in pressure intervals), 
where nuclear transformations take place.
In the GC models,  they occur at some values of $P<P_\mathrm{oi}$, at
$P=P_\mathrm{oi}$, and at $P$ close to the bottom of the inner crust
(where nuclei completely disintegrate into neutrons due to instability discussed in GC).
The latter heat almost entirely leaks into the core because of the steep
increase of thermal conductivity in the crust. The
total heat produced in the outer and inner crust per each accreted
baryon, $E_\mathrm{h}$, increases with an increase of $P_\mathrm{oi}$,
while the heat produced at
the outer-inner crust interface decreases (because the
reactions there become less efficient at a larger electron Fermi momentum,
corresponding to a larger $P_\mathrm{oi}$). The pressure
values, at which the nuclear transformations occur, as well as the
values of $P_\mathrm{oi}^\mathrm{(0)}$, $P_\mathrm{oi}^\mathrm{(cat)}$
and the amounts of heat released at each transformation, depend on the
employed nuclear models. \citet{GusakovChugunov21} presented these
values for the finite-range droplet macroscopic model FRDM12
\citep{Moeller_16} and  the family of models based on
Hartree-Fock-Bogoliubov calculations with Skyrme-like effective nuclear
forces BSk24, BSk25, BSk26 \citep*{GorielyCP16}.

\begin{figure}
\centering
\includegraphics[width=\columnwidth]{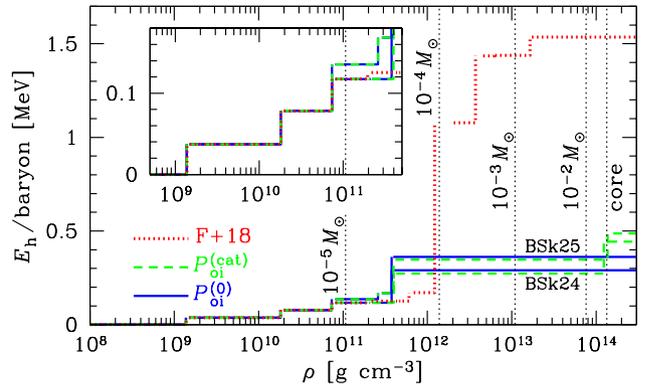}
\caption{Total heat, $E_\mathrm{h}$, generated per accreted baryon 
in a layer from the star surface to a given density, as a function of mass
density $\rho$, according to the model of \citet{Fantina_18} (F+18,
dotted red line) and models of \citet{GusakovChugunov21} with
$P_\mathrm{oi}=P_\mathrm{oi}^\mathrm{(cat)}$ (dashed green lines) and
$P_\mathrm{oi}=P_\mathrm{oi}^\mathrm{(0)}$ (solid blue lines), based on
the models of nuclear interactions consistent with the EoSs BSk24 (lower
lines) and BSk25 (upper lines). The gaps on the lines (best
visible in the F+18 case) are due to the density jumps at the interfaces
between neighbouring layers containing different nuclei. The  right vertical
dotted line corresponds to the crust-core transition according to the
EoS BSk24; the other vertical dotted lines mark the $\rho$ values
corresponding to four masses of overlying material, from
$10^{-5}\,\msun$ to $10^{-2}\,\msun$, labelled near these lines, for
a neutron star with gravitational mass $M=1.4\,\msun$ and radius
$R=12.6$~km. The inset shows a zoom-in of the low-density region.
}
\label{fig:heatsrc}
\end{figure}

Fig.~\ref{fig:heatsrc} shows the total heat generated per accreted
baryon, from the surface to a given density in the crust, as a function
of mass density, for the F+18 model of the accreted crust and the GC
models based on the BSk24 and BSk25 nuclear forces (which provide more
plausible mass thresholds for rapid neutron-star cooling than BSk22 and
BSk26, as argued by \citealt{Pearson_18}).  We use the version of the
F+18 model based on the BSk21 nuclear forces \citep*{GorielyCP10} (table
A.1 of \citealt{Fantina_18}). For the GC models, we consider the minimum
and maximum values of $P_\mathrm{oi}$. The corresponding crust structure
and heat production are presented in Tables~\ref{tab:GCheat24} and
\ref{tab:GCheat25}. We assumed that the  number fractions of protons
$Y_\mathrm{p}=Z/A'$ and free neutrons $Y_\mathrm{nf}=1-A/A'$ among all
nucleons equal
the $Y_\mathrm{p}$ and $Y_\mathrm{nf}$ values in the non-accreted crust,
fitted by \citet{Pearson_18}.
Here $Z$, $A$ and $A'$ are the numbers of protons, bound
nucleons and all nucleons in a Wigner-Seitz cell, respectively
(they are needed to calculate the ion heat capacity and
electron-ion transport coefficients). We also assume that the residual
heat release in the deep layers of the inner crust, predicted by
\citet{GusakovChugunov21}, occurs at the proton-drip point determined by
\citet{Pearson_18}.

\begin{table}
\caption{Nuclear composition and heat release in the accreted
neutron-star crust according to the GC models
\citep{GusakovChugunov20,GusakovChugunov21} using the BSk24 nuclear mass
table. The columns show the pressure $P$, at which a nuclear reaction
occurs, the mass density $\rho$ before  and after the reaction, the
corresponding nuclear charge and mass numbers $Z$ and $A$
 and the heat release per an accreted nucleon $\Delta
E_\mathrm{h}$. The upper part of the table describes the outer crust
and does not depend on $P_\mathrm{oi}^\mathrm{(0)}$, the middle part
describes the interface at $P=P_\mathrm{oi}$ for 
$P_\mathrm{oi}=P_\mathrm{oi}^\mathrm{(0)}$ and the bottom part
corresponds to the case of $P_\mathrm{oi}=P_\mathrm{oi}^\mathrm{(cat)}$.
The last line shows the heat produced near the crust bottom; in this
case we use the $\rho$ and $P$ values and the ratio $Z/A$
at the proton drip (not accompanied by a sharp nuclear transformation)
according to \citet{Pearson_18}.
}
\label{tab:GCheat24}
\begin{tabular}{c @{\hspace{3ex}} c  @{\hspace{3ex}} c  @{\hspace{3ex}} c  @{\hspace{3ex}} c}
\hline
$P$ [dyn cm$^{-2}$] & $\rho$ [\gcc] & $Z$  & $A$ & $\Delta E_\mathrm{h}$ [keV]  \\
\hline
$6.46\times10^{26}$  & (1.37--1.48)$\times 10^9 $ & $26\to24$   & $56$ & 37 \\
$1.83\times10^{28}$  & (1.81--1.97)$\times 10^{10} $ & $24\to22$   & $56$ & 41 \\
$1.06\times10^{29}$  & (7.36--8.08)$\times 10^{10} $ & $22\to20$   & $56$ & 39 \\
$3.36\times10^{29}$  & (1.93--2.03)$\times 10^{11} $ & $20\to19$   & $56$ & 0 \\
$3.50\times10^{29}$  & (2.09--2.20)$\times 10^{11} $ & $19\to18$   & $56$ & 0 \\
\hline
\multicolumn{5}{c}{$P_\mathrm{oi}=P_\mathrm{oi}^\mathrm{(0)}$}\\
$7.16\times10^{29}$  & (3.77--3.98)$\times 10^{11} $ & $18\to20$   & $56\to64$ & $172$ \\
\hline
\multicolumn{5}{c}{\mbox{$P_\mathrm{oi}=P_\mathrm{oi}^\mathrm{(cat)}$}}\\
$7.73\times10^{29}$  & (3.99--4.28)$\times 10^{11} $ & $18\to20$   & $56\to65$ & $156$ \\
$3.72\times10^{32}$  & $1.23\times 10^{14} $ & $20$   & $92$ & $169$ \\
\hline
\end{tabular}
\end{table}

\begin{table}
\caption{The same as in Table~\ref{tab:GCheat24} but for the BSk25
nuclear masses.
}
\label{tab:GCheat25}
\begin{tabular}{c @{\hspace{3ex}} c  @{\hspace{3ex}} c  @{\hspace{3ex}} c  @{\hspace{3ex}} c}
\hline
$P$ [dyn cm$^{-2}$] & $\rho$ [\gcc] & $Z$  & $A$ & $\Delta E_\mathrm{h}$ [keV]  \\
\hline
$6.46\times10^{26}$  & (1.37--1.48)$\times 10^9 $ & $26\to24$   & $56$ & 37 \\
$1.83\times10^{28}$  & (1.81--1.97)$\times 10^{10} $ & $24\to22$   & $56$ & 41 \\
$1.06\times10^{29}$  & (7.36--8.08)$\times 10^{10} $ & $22\to20$   & $56$ & 39 \\
$5.00\times10^{29}$  & (2.59--2.88)$\times 10^{11} $ & $20\to18$   & $56$ & 33 \\
\hline
\multicolumn{5}{c}{$P_\mathrm{oi}=P_\mathrm{oi}^\mathrm{(0)}$}\\
$7.16\times10^{29}$  & (3.77--3.98)$\times 10^{11} $ & $18\to20$   & $56\to64$ & $193$ \\
\hline
\multicolumn{5}{c}{\mbox{$P_\mathrm{oi}=P_\mathrm{oi}^\mathrm{(cat)}$}}\\
$7.63\times10^{29}$  & (3.95--4.18)$\times 10^{11} $ & $18\to20$   &
$56\to64$ & $156$ \\
$3.33\times10^{32}$  & $1.37\times 10^{14} $ & $20$   & $112$ & $87$ \\
\hline
\end{tabular}
\end{table}

The total heat release $E_\mathrm{h}$ in the entire accreted crust
varies from 0.29 MeV for  $P_\mathrm{oi}=P_\mathrm{oi}^\mathrm{(0)}$ to
0.44 MeV for $P_\mathrm{oi}=P_\mathrm{oi}^\mathrm{(cat)}$ in the BSk24
case (Table~\ref{tab:GCheat24}) and from 0.36 MeV for
$P_\mathrm{oi}=P_\mathrm{oi}^\mathrm{(0)}$ to 0.49 MeV for
$P_\mathrm{oi}=P_\mathrm{oi}^\mathrm{(cat)}$ in the BSk25 case
(Table~\ref{tab:GCheat25}). These values are several times smaller than
traditional ones: for example, $E_\mathrm{h}=1.54$ MeV in the F+18 case.

\section{Equilibrium thermal emission}
\label{sect:long}

If the post-outburst thermal relaxation of the crust lasts sufficiently
long time, the crust approaches thermal equilibrium with the core. 
Then the luminosity  is a function of the core
temperature, which is determined by the balance
between energy income due to the accretion and the energy
losses due to neutrino and photon emission. Since the
time needed for an appreciable heating or cooling of the core is much
longer than the accretion variability \citep{Colpi_01}, the
equilibrium level is a function of the \emph{average mass accretion
rate} $\langle\dot{M}\rangle$. Here and hereafter, the angle brackets
$\langle\ldots\rangle$ denote averaging over a time-span covering many
outburst and quiescence cycles. The dependence of the equilibrium
luminosity on $\langle\dot{M}\rangle$ is called \emph{heating curve}
\citep*{YakovlevLH03}.

Since the mass of a neutron-star crust is small ($\sim10^{-2}\,\msun$),
the initial ground-state crust can be completely replaced by the
reprocessed accreted material, if the accretion lasts long enough. The
short spin periods of neutron stars in the SXTs ($<10$ ms for all but
one SXTs with measured spin periods; see, e.g., table~2 in
\citealt{PotekhinCC19} and references therein) can be explained by
recycling due to the accretion of an appreciable mass.
Therefore, in this paper we restrict ourselves to the fully accreted
crust models.\footnote{Partially accreted crusts of neutron stars in
SXTs have been discussed
in the frames of the traditional approach by \citet*{WijnandsDP13};
\citet{Fantina_18,PotekhinCC19,Suleiman_22}.}

We performed calculations of quiescent luminosities of neutron stars in
the SXTs, similar to those presented in \citet{PotekhinCC19}, with
essentially the same basic physics input. For the core
EoS we consider the models BSk24, BSk25
\citep{Pearson_18} and APR$^*$ (\citealt{APR}, in
the parametrized form of \citealt{PotekhinChabrier18}). The heat
capacities, neutrino energy losses and thermal conductivities are
treated essentially in the same way as described by
\citet{PotekhinChabrier18}, except for a small improvement for the
electron thermal conductivities of outer envelopes composed of helium,
described below. The employed models of baryon superfluidity are
discussed
in Section~\ref{sect:superflu}. 
 
\begin{figure}
\includegraphics[width=\columnwidth]{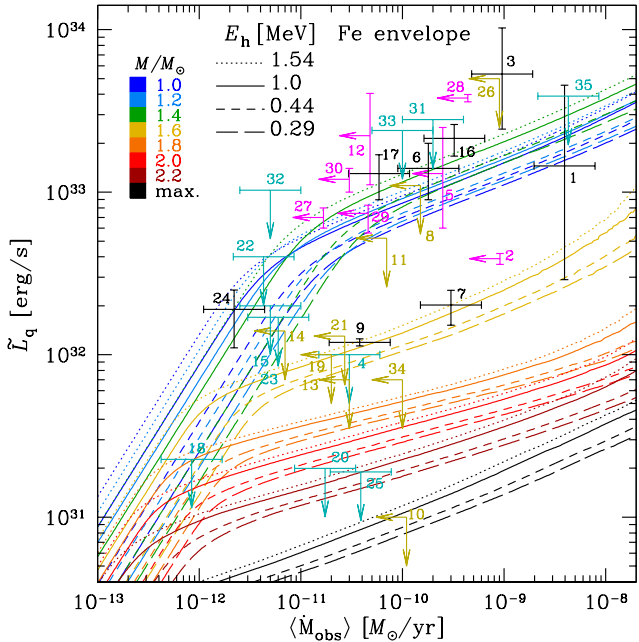}
\caption{Quiescent thermal luminosities of SXTs as functions of average
accretion rates in the reference frame of a distant observer. The iron
heat-blanketing envelope is assumed. Solid errorbars and arrows show the
data collected in \citet{PotekhinCC19} for the observed SXTs coded with
numbers: 4U 2129+47 (1), KS 1731$-$260 (2), 4U 1608$-$522 (3), EXO
1745$-$248 (4), 1M1716$-$315 (5), XTE 1709.5$-$267 (6), MXB 1659$-$29
(7), XB 1732$-$304 (8), Cen X-4 (9), 1H 1905+000 (10), SAX
J1806.8$-$2435 (11), 4U 1730$-$22 (12), EXO 1747$-$214 (13), XTE
2123$-$058 (14), Aql X-1 (15), 4U1908+005 (16), SAX J1748.9$-$2021 (17),
NGC 6440 X-2 (18), XTE J0929$-$314 (19), SAX J1808.4$-$3658 (20), XTE
J1807$-$294 (21), XTE J1751$-$305 (22), XTE J1814$-$338 (23), IGR
J00291+5934 (24), HETE J1900.1$-$2455 (25), XTE J1701$-$462 (26), IGR
J17480$-$2446 (27), EXO 0748$-$676 (28), 1RXS J180408.9$-$342058 (29),
Swift J174805.3$-$244637 (30), SAX J1750.8$-$2900 (31), Swift
J1756.9$-$2508 (32), GRS 1747$-$312 (33), IGR J18245$-$2452 (34) and
MAXI J0556$-$332 (35). The data for objects 7, 25, 32 and 35 
have been updated, as explained in the text. 
The lines show theoretical predictions for neutron stars of
different masses from $1.0\,\msun$ to the maximum mass for the BSk24 EoS
($M_\mathrm{max}=2.28\,\msun$), 
according to the colour map, computed for the heating power
predicted either by the model F+18 (dotted lines) or by the
thermodynamically consistent GC models with
$P_\mathrm{oi}=P_\mathrm{oi}^\mathrm{(0)}$ (long-dashed lines) and
$P_\mathrm{oi}=P_\mathrm{oi}^\mathrm{(cat)}$ (short-dashed lines). The
solid lines correspond to the net neutron-star heating fixed at
$E_\mathrm{h}=1.0$ MeV per accreted baryon. 
}
\label{fig:qemxsf25gs}
\end{figure}

Fig.~\ref{fig:qemxsf25gs} shows the heating curves for neutron stars of
different masses, based on the EoS model BSk24. The long-term average
accretion rates $\langle \dot{M}\rangle$ and quasi-equilibrium
bolometric thermal luminosities in quiescence $L_\mathrm{q}$ are plotted
taking account of the gravitational redshift, as seen by a distant
observer (thus corrected, they are denoted $\langle
\dot{M}_\mathrm{obs}\rangle$ and $\tilde{L}_\mathrm{q}$; see
\citealt{PotekhinCC19} for their calculation). The dashed heating curves
are calculated using the GC crust models with the minimum and maximum
values of the pressure $P_\mathrm{oi}$ at the outer-inner crust
interface, as described in Section~\ref{sect:accrust}, while the dotted
heating curves are computed according to the F+18 model. The curves of
different colours correspond to neutron-star models with different
masses. The error bars (arrows) represent the estimates of (upper limits
to) $\langle \dot{M}_\mathrm{obs}\rangle$ and $\tilde{L}_\mathrm{q}$,
derived from observations of the SXTs, as listed in \citet{PotekhinCC19}
but with updates for several objects: MXB 1659$-$29, HETE
J1900.1$-$2455, Swift
J1756.9$-$2508 and MAXI J0556$-$332.
For MXB 1659$-$29 (object 7), an improved estimate
of $\langle \dot{M}_\mathrm{obs}\rangle\sim 3\times10^{-10}\,\msun$
yr$^{-1}$ \citep{PotekhinChabrier21} is used. For HETE J1900.1$-$2455
(object 25), an updated estimate of
$\tilde{L}_\mathrm{q}\approx\mbox{(1--2)}\times10^{31}$ erg~s$^{-1}$
follows from the results obtained by \citet{Degenaar_21} with
atmospheric fits to the spectrum measured in 2018. Note that the crust
of the neutron star in HETE J1900.1$-$2455 may not have yet reached
thermal quasi-equilibrium by 2018 \citep{Degenaar_21}, therefore we
consider this result as an upper
limit. For Swift
J1756.9$-$2508 (object 32), we updated the estimate of the average
accretion rate, $\langle \dot{M}_\mathrm{obs}\rangle = 5
\times10^{-12}\msun\mbox{ yr}^{-1}$, according to \citet{Li_21}.

For MAXI J0556$-$332 (object 35), we refined the data using the results
recently published by \citet{Page_22}. Four outbursts have been observed
since its discovery in 2011, and the crust may not have reached thermal
equilibrium between them. Therefore, just as in the case of HETE
J1900.1$-$2455, the measured quiescent thermal luminosities can only be
treated as upper limits to the quasi-equilibrium luminosity of MAXI
J0556$-$332. The strongest upper limit $\tilde{L}_\mathrm{q}\lesssim
4\times10^{33}$ erg~s$^{-1}$ is provided by the results of the analysis
of the \textit{XMM-Newton} observation taken in 2019 (section 2.3 in
\citealt{Page_22}); it refines the previous upper bound. We have also
refined the estimate of $\langle \dot{M}_\mathrm{obs}\rangle$ for this
source. Integrating $\dot{M}$ presented in the top panel of fig.~6 in
\citet{Page_22} over time and dividing the result by the total time of
observations, we obtain $\langle
\dot{M}_\mathrm{obs}\rangle\approx4.3\times10^{-9}\,\msun\mbox{~yr}^{-1}$.

The estimates of $\langle \dot{M}_\mathrm{obs}\rangle$ are rather
uncertain, because the history of the SXTs observations (up to a few
decades) is much shorter than the time of neutron star core heating due
to accretion, which would be a proper interval for the time averaging.
The estimate by \citet{WijnandsDP13} for the latter time scale can be
written as $\tau_\mathrm{th}\approx\mbox{(0.3--3)}\times 10^4\,
L_{32}\,(\dot{M}_{-11}E_\mathrm{h1})^{-1}$ yr, where
$L_{32}\equiv L_\mathrm{q}/(10^{32}$ erg~s$^{-1}$),
$\dot{M}_{-11}\equiv \langle\dot{M}\rangle/(10^{-11}\,\msun\mbox{
yr}^{-1})$ and $E_\mathrm{h1} \equiv E_\mathrm{h}/1$~MeV.
Therefore, the currently observed average  accretion rate might be
non-representative for some X-ray transients (see, e.g.,
\citealt{WijnandsDP13} for a discussion). In Fig.~\ref{fig:qemxsf25gs}
we conditionally plot uncertainties of $\langle
\dot{M}_\mathrm{obs}\rangle$ as a factor 2 around the most likely values
based on the available observations (in the case of MXB 1659$-$29 this
factor somewhat underestimates the true uncertainty -- see
\citealt{PotekhinChabrier21}).

\begin{figure}
\includegraphics[width=\columnwidth]{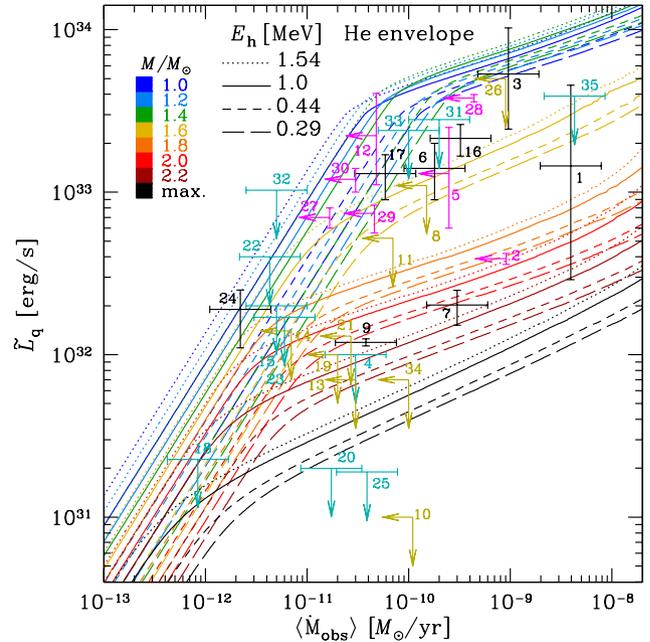}
\caption{The same as in Fig.~\ref{fig:qemxsf25gs}, but with the
heat-blanketing  envelope composed of light elements (`He envelope', see
text).
}
\label{fig:qemxsf25He}
\end{figure}

In Fig.~\ref{fig:qemxsf25gs}, the outer crust of the star is assumed to
be composed of the elements listed in Table~\ref{tab:GCheat24} at all
densities. In this case, the heat-blanketing envelope  is made of iron
within the `sensitivity strip' \citep*{GPE83} at $\rho\lesssim10^9$
\gcc.
Such a heat blanket is
a relatively poor heat conductor. As the
opposite extreme, in Fig.~\ref{fig:qemxsf25He} we show the analogous
heating curves calculated assuming that the heat blanket is made of the
largest sustainable amount of helium, with carbon and oxygen beneath the
helium layer. Such an envelope is most transparent to the heat transport
\citep*[see][for a review]{BeznogovPY21}; we call it the He envelope for
short. We have taken account of the correction by \citet{Blouin_20}
for the helium thermal conductivity at moderate electron degeneracy,
modified according to \citet{Cassisi_21} (we used the modification
version dubbed `B20sd' by the latter authors). This update slightly
affects the long-term cooling and heating curves of neutron
stars with envelopes composed of light elements.

\subsection{The long-term effect of shallow heating}

From Figs.~\ref{fig:qemxsf25gs} and~\ref{fig:qemxsf25He} we see that it
is possible to reach a satisfactory agreement between the theory and
observations for each SXT by varying the neutron star mass and envelope
composition. However, for the hottest objects at the given accretion
rates, such as IGR J00291+5934 (object 24), IGR J17480$-$2446 (object
27) or Swift J174805.3$-$244637 (object 30), the thermodynamically
consistent GC models give somewhat larger discrepancies with
observations than the F+18 model. This tension between the theory and
observations can be alleviated by including the shallow heating, that is
an additional heating at relatively low densities $\rho\sim10^8-10^{10}$
\gcc{}, beside the predictions of the deep crustal heating theory. The
shallow heating has been introduced into the theory by
\citet{BrownCumming09} to match the slope between the first two
observations of the SXT MXB 1659$-$29 after it had turned into the
quiescent state. Later the shallow heating  proved to be necessary to
fit the theory with observations for  most of the SXTs with observed
crustal cooling in quiescence (see, e.g., results and discussions in
\citealt{Deibel_15,Waterhouse_16,WijnandsDP17}). Different theoretical
fits to observed crust cooling curves  in different SXTs require shallow
heat deposit $E_\mathrm{sh}$ from 0 to several MeV per accreted nucleon,
most typical values being $E_\mathrm{sh}\sim(0.2-2)$ MeV per nucleon
(see \citealt{Chamel_20} for a summary and references; see also
Section~\ref{sect:MXB} below). 

\begin{figure}
\centering
\includegraphics[width=\columnwidth]{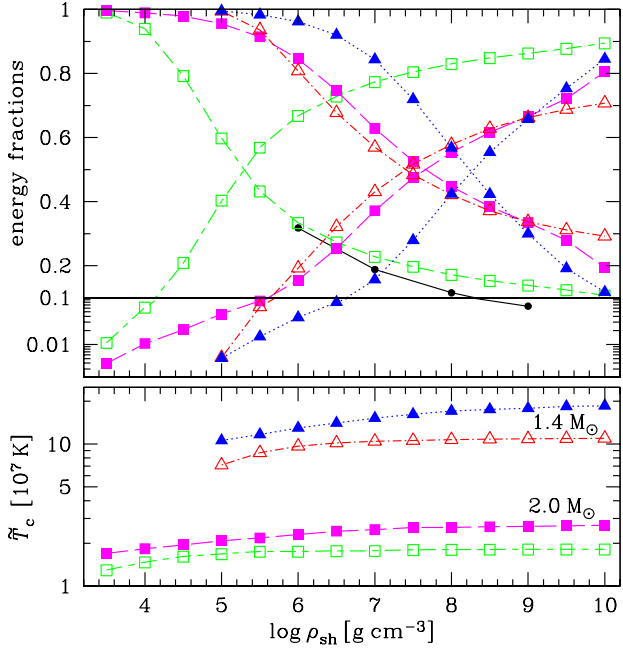}
\caption{\textit{Top panel}: The fractions of thermal energy deposited 
in a narrow layer close to  a density $\rho_\mathrm{sh}$ leaking through
the surface (decreasing dependences) and flowing into the core
(increasing ones), computed assuming a stationary accretion. Triangles
are for a neutron star with mass $M=1.4\,\msun$ and radius $R=12.6$ km
with the iron heat-blanketing envelope; squares are for a neutron star
with $M=2.0\,\msun$, $R=12.4$ km and the He envelope; empty symbols
correspond to an accretion with
$E_\mathrm{sh}\times\dot{M}=1\mbox{~MeV}\times10^{-10}\,\msun$ yr$^{-1}$
(then the total heating power is $H\approx6\times10^{33}$ erg s$^{-1}$
in the local reference frame); filled symbols correspond to a higher
accretion rate
$E_\mathrm{sh}\times\dot{M}=1\mbox{~MeV}\times10^{-8}\,\msun$ yr$^{-1}$
($H\approx6\times10^{35}$ erg s$^{-1}$). The lines connecting the
symbols serve as a guide for the eye. For comparison, the dots
connected by the solid line show the fraction of energy from a short
($\sim2$ minutes)  superburst that is radiated from the atmosphere,
according to the simulation results of \citet{Yakovlev_21}. In the lower
part of this panel, we use the logarithmic scale of the energy fraction
to enhance visibility of its smallest values. \textit{Bottom panel}: The
stationary redshifted temperature in the core (in units of $10^7$~K)
provided by the respective heating models shown using the same symbols
and line styles as in the top panel.
}
\label{fig:fluxfrac}
\end{figure}

\begin{figure}
\centering
\includegraphics[width=\columnwidth]{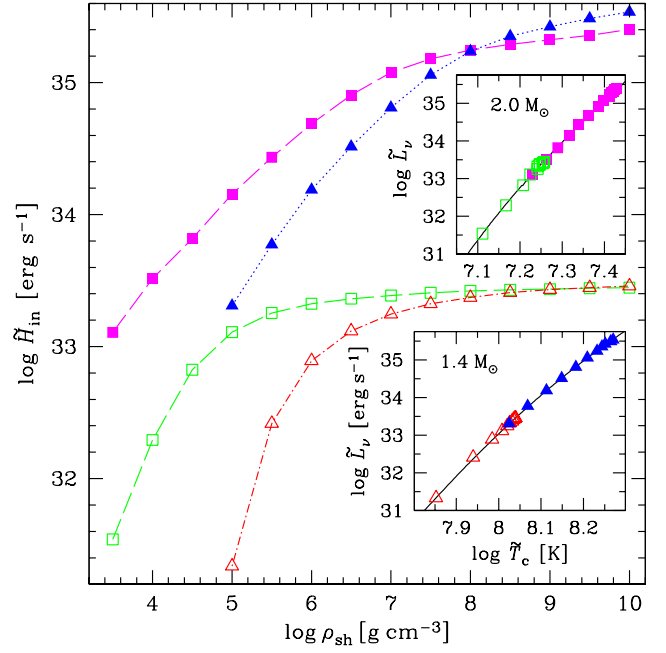}
\caption{Logarithm of the redshifted heating power
$\tilde{H}_\mathrm{in}$ injected into the core as function of
$\rho_\mathrm{sh}$ for the same neutron star and heating models as in
Fig.~\ref{fig:fluxfrac}, plotted with the same symbols and line styles.
The insets show logarithms of the total redshifted neutrino luminosity
$\tilde{L}_\nu$ as a function of redshifted temperature in the core
($\tilde{L}_\nu=\tilde{H}_\mathrm{in}$ in the stationary state).
}
\label{fig:fluxd}
\end{figure}

\citet*{OotesWP19} noted that a large part of the  shallow heat 
$E_\mathrm{sh}$ flows inside the star (we will denote this part
$E_\mathrm{sh}^\mathrm{in}$ hereafter) and significantly contributes to
the equilibrium state, resulting in an increase of the quiescent
luminosity.  Our calculations support this conclusion. To test it, we
have numerically simulated thermal evolution of a neutron star assuming
the shallow heating power concentrated at mass densities around
$\rho_\mathrm{sh}$ in the narrow interval
$\rho\in(0.95\rho_\mathrm{sh},1.05\rho_\mathrm{sh})$ according to the
Wigner semicircle distribution. In the simulations we fixed
$E_\mathrm{sh}=1$ MeV per an accreted nucleon and varied the accretion
rate, the star mass and radius, and the envelope composition. Some
results are presented in Figs.~\ref{fig:fluxfrac} and~\ref{fig:fluxd}.
The top panel in Fig.~\ref{fig:fluxfrac} shows the heating energy
fractions that are either absorbed in the core  or radiated from the
surface as functions of the heating density $\rho_\mathrm{sh}$ for
neutron-star models with $M=1.4\,\msun$ and $M=2.0\,\msun$, constructed
according to the EoS BSk24, with the iron heat-blanketing envelope for
the lighter star and the He envelope for the heavier star model.  
%
%
The energy fraction that eventually flows into the core,
$E_\mathrm{sh}^\mathrm{in}/E_\mathrm{sh}$, depends on the model,  but
in any case we have
$E_\mathrm{sh}^\mathrm{in}/E_\mathrm{sh}\gtrsim1/2$ at 
$\rho_\mathrm{sh}\gtrsim10^8$ \gcc. 
The bottom panel of
Fig.~\ref{fig:fluxfrac} shows the temperature supported by the the
stationary shallow heating (without other heating sources) in the deep
layers of the inner crust and outer layers of the core. The inner
temperatures are lower for the heavier neutron star model than for the
lighter one because of the enhanced neutrino emission due to the direct
Urca processes. Fig.~\ref{fig:fluxd} shows the
$\rho_\mathrm{sh}$-dependence of the heating power supplied to the
core,  ${H}_\mathrm{in}=E_\mathrm{sh}^\mathrm{in}\dot{M}/m_\mathrm{n}$,
where $m_\mathrm{n}$ is the nucleon mass. The insets illustrate the
steep increase of the star's neutrino luminosity with increasing
temperature of the core, which explains the weakness of the temperature
variation with changing $\rho_\mathrm{sh}$ in the bottom panel of
Fig.~\ref{fig:fluxfrac}. The plotted quantities are redshifted (i.e.,
reduced to the distant observer's reference frame).

We have neglected the energy produced by the impact of accreted
particles on the surface of the star. Some fraction of this energy
propagates to the deeper layers and heats up the interior. 
The typical
gravitational energy release of $\sim200$ MeV per accreted baryon could
provide an appreciable internal heating, if the fraction of the surface
heat reaching the interior
were $\gtrsim10^{-3}$. 
However, $E_\mathrm{sh}^\mathrm{in}/E_\mathrm{sh}$ decreases so quickly
as the heating location approaches the surface (cf.\
Fig.~\ref{fig:fluxfrac}) that the required energy fraction
$\gtrsim 10^{-3}$ seems unlikely.

The surface heating may also affect the outer boundary conditions for
the heat diffusion problem through the dependence of thermal
conductivity on temperature. This effect is hard to quantify, because
the distribution of the surface heat under the spreading layer on the
neutron star surface is poorly known, in particular because of its
variability in time and space (see, e.g., \citealt{GilfanovSunyaev14} and
references therein). Anyway, the model of stationary accretion
that we employ in this section is only a crude approximation.
Simulations of the short-term thermal response of a neutron star to a
burst in the outer crust give qualitatively similar, but quantitatively
different dependence of the heat fraction eventually radiated through
the surface (the dots connected by solid lines in the top panel of
Fig.~\ref{fig:fluxfrac}, after \citealt{Yakovlev_21}).

\citet{Chamel_20} have shown that carbon or oxygen fusion, followed by
electron captures in the crust of accreting neutron stars could provide
up to $\approx1.4$ MeV per an accreted nucleon (with the reservation
that the actual energy release depends on the abundance of these
elements). Then it would be
plausible that up to $\sim 1$ MeV per accreted nucleon could be
supplied to the core due to the shallow heating. For illustration, in
Figs.~\ref{fig:qemxsf25gs} and~\ref{fig:qemxsf25He} we plot the heating
curves with $E_\mathrm{h}=1$~MeV, which corresponds to
$E_\mathrm{sh}^\mathrm{in}$ equal to 0.71 MeV and 0.56 MeV in the
cases where $P_\mathrm{oi}=P_\mathrm{oi}^\mathrm{(0)}$ and
$P_\mathrm{oi}=P_\mathrm{oi}^\mathrm{(cat)}$, respectively.

\begin{figure}
\includegraphics[width=\columnwidth]{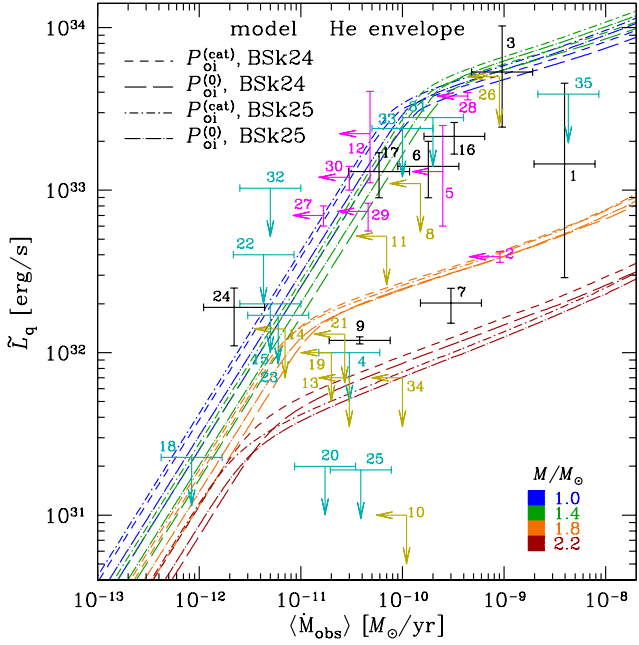}
\caption{The observational points and heating curves from
Fig.~\ref{fig:qemxsf25He} for $M=1.0\,\msun$, 1.4\,\msun, 1.8\,\msun{} and
2.2\,\msun{}  for the EoS  BSk24 and the GC heating model according to
Table~\ref{tab:GCheat24} (long and
short dashed lines for $P_\mathrm{oi}=P_\mathrm{oi}^\mathrm{(0)}$ and
$P_\mathrm{oi}=P_\mathrm{oi}^\mathrm{(cat)}$) compared with the
analogous heating curves for the EoS BSk25 and the heating power
according to Table~\ref{tab:GCheat25} (dot-long-dash and dot-short-dash
lines for $P_\mathrm{oi}=P_\mathrm{oi}^\mathrm{(0)}$ and
$P_\mathrm{oi}=P_\mathrm{oi}^\mathrm{(cat)}$, respectively).
}
\label{fig:qemxBSk24BSk25}
\end{figure}

\subsection{The effects of the EoS choice}

The EoS model BSk25 is somewhat stiffer than BSk24 and provides somewhat
stronger deep crustal heating. We find that the use of the BSk25 model
for the heating profile and crust composition
(Table~\ref{tab:GCheat25}), as well as for the EoS, effective nucleon
masses and composition of the core \citep{Pearson_18} gives very similar
results to the case of the BSk24 model. Fig.~\ref{fig:qemxBSk24BSk25}
presents a comparison of the heating curves obtained with these two
models for the GC crust with the smallest and largest values of the
parameter $P_\mathrm{oi}$.

The faintness of the coldest neutron stars in SXTs at quiescence, such
as objects 4, 7, 9, 20, 25 in Figs.~\ref{fig:qemxsf25gs},
\ref{fig:qemxsf25He}, \ref{fig:qemxBSk24BSk25}, can be explained by
neutrino emission from their interiors due to powerful direct Urca
processes. Such processes are only allowed at sufficiently large proton
fraction $Y_\mathrm{p}$ ($Y_\mathrm{p}>1/9$ in the simplest case of the
$npe$ matter -- \citealp[see, e.g.,][]{Haensel95}). This rules out the
theoretical models that predict smaller $Y_\mathrm{p}$ at any densities
relevant to the cores of neutron stars. Since $Y_\mathrm{p}$ usually
increases with density in the core, the direct Urca processes are
allowed at densities exceeding a certain threshold,
$\rho>\rho_\mathrm{DU}$, which can be attained only in the central parts
of neutron stars with masses above some threshold $M_\mathrm{DU}$. For
the BSk24 and BSk25 models, $\rho_\mathrm{DU}=8.25\times10^{14}$ \gcc{}
and $8.56\times10^{14}$ \gcc, respectively. In both cases
$M_\mathrm{DU}\approx1.6\,\msun$ \citep{Pearson_18}. For this reason,
the relatively faint objects in Figs.~\ref{fig:qemxsf25gs},
\ref{fig:qemxsf25He}, \ref{fig:qemxBSk24BSk25} are matched by the
heating curves for $M > 1.6\,\msun$. 

\begin{figure}
\includegraphics[width=\columnwidth]{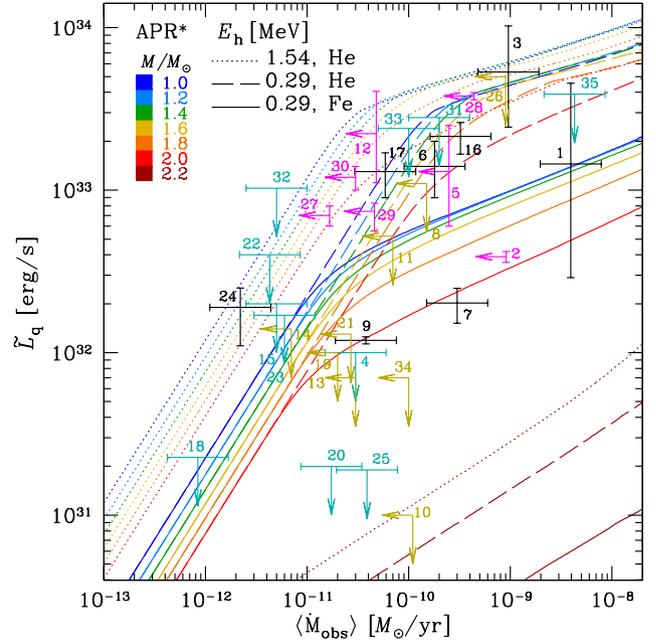}
\caption{The observational points and heating curves for neutron stars
of different masses and the GC model for
$P_\mathrm{oi}=P_\mathrm{oi}^\mathrm{(0)}$ with the He envelope
(long-dashed lines) and heavy-element (Fe) envelope (solid curves), as
in Figs.~\ref{fig:qemxsf25gs} and~\ref{fig:qemxsf25He},  but computed
using the APR$^*$ model for the EoS and composition of the core.
}
\label{fig:qemxAPR}
\end{figure}

Fig.~\ref{fig:qemxAPR} demonstrates the heating
curves computed using the widely used EoS model APR$^*$ \citep*{APR} in
the neutron star core (as parametrized by \citealt{PotekhinChabrier18})
and the BSk24 model in the crust. In this case,
$\rho_\mathrm{DU}=1.56\times10^{15}$ \gcc{} and
$M_\mathrm{DU}=2.006\,\msun$, which necessitates fine tuning of the mass
around $M\approx2\,\msun$ to match the heating
curves of some relatively faint SXTs with the observations.
However, still
much finer tuning would be needed, if we neglected the enhancement of the modified Urca
processes (which are the main neutrino emission processes at
$\rho<\rho_\mathrm{DU}$) at $\rho\sim\rho_\mathrm{DU}$, discovered by \citet{ShterninBH18}. This
enhancement becomes very strong at $\rho\approx\rho_\mathrm{DU}$, which
may be qualitatively regarded as `smearing' the direct Urca threshold.
For the APR$^*$ model, this effect noticeably enhances the total
neutrino luminosity of the neutron stars with $M\approx(1.8-2.0)\,\msun$
and thus decreases their temperatures. In the absence of such
enhancement, the heating curves of all neutron stars with $M\leq2\,\msun$
in Fig.~\ref{fig:qemxAPR} would be close to each other.

\subsection{The effects of baryon superfluidity}
\label{sect:superflu}

\begin{figure}
\centering
\includegraphics[width=\columnwidth]{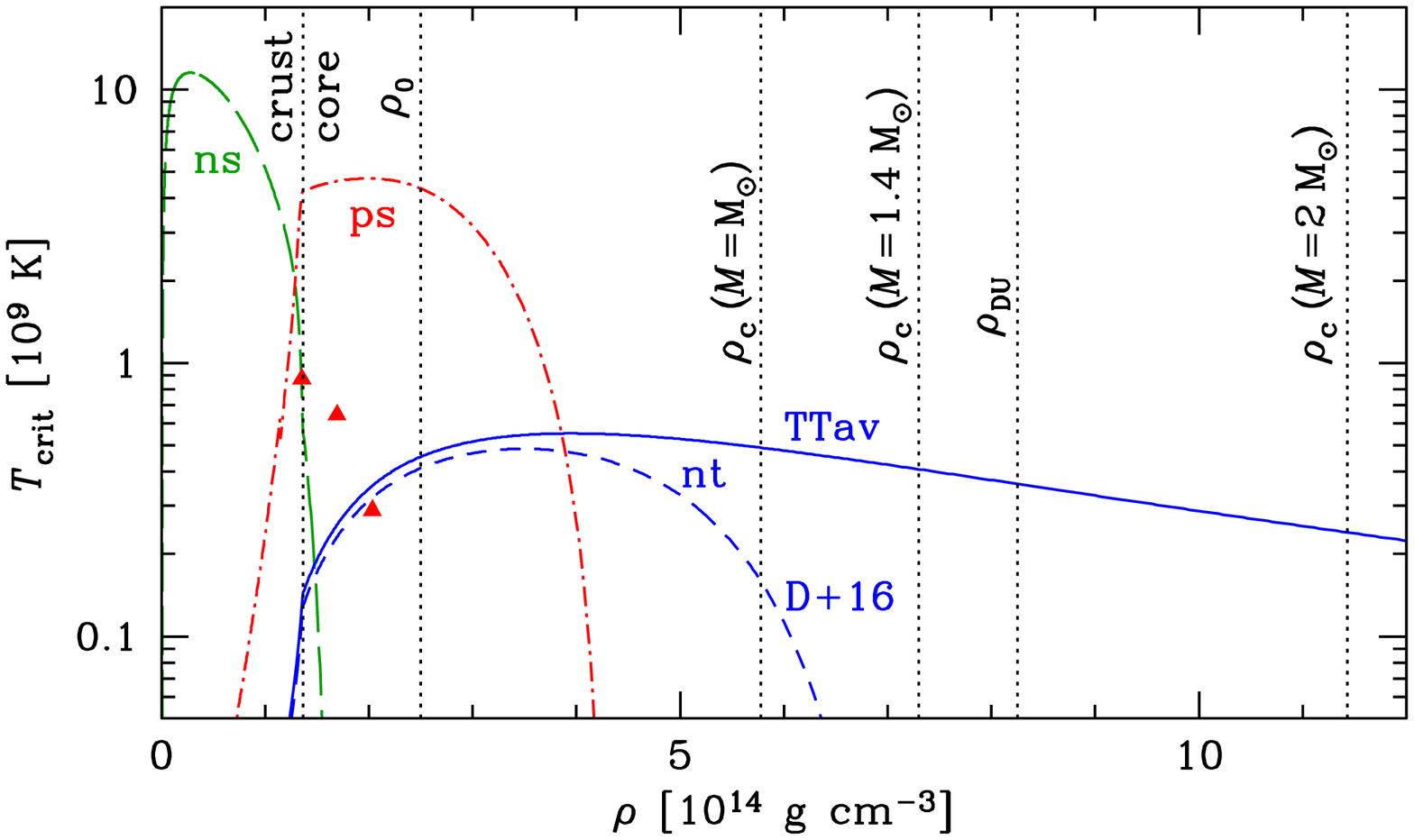}
\caption{Critical temperature for nucleon superfluidity as a function of
mass density for neutron singlet (ns), proton singlet (ps) and neutron
triplet (nt) pairing. The long-dashed, dot-dash and solid lines
correspond to the superfluid gap parametrizations by \citet{Ho_15}
named MSH, BS and TTav for theoretical calculations by
\citet{MargueronSH08} for the ns pairing,
\citet{BaldoSchulze07} for the ps pairing and
\citet{TakatsukaTamagaki04} for the nt pairing,
respectively. The short-dashed line shows the parametrization 
by \citet{Ding_16} to their calculations for the nt
pairing, and the triangles show the results of \citet{Guo_19} for
the ps pairing. The vertical dotted lines are drawn at the densities
corresponding to the crust-core boundary, nuclear saturation
density $\rho_0$ \citep*{HorowitzPR20}, direct Urca threshold $\rho_\mathrm{DU}$
and densities $\rho_\mathrm{c}$ at the centre 
of neutron stars with masses $M=1\,\msun$, $1.4\,\msun$ and $2.0\,\msun$,
according to the EoS BSk24 \citep{Pearson_18}.
}
\label{fig:tcrit}
\end{figure}

The thermal evolution of neutron stars can be affected by baryon
superfluidity (see, e.g., \citealt{Page_14,SedrakianClark19}, for review
and references). The three principal types of the superfluidity arise
from neutron singlet (ns), proton singlet (ps) and neutron triplet (nt)
types of Cooper pairing. The effects of each superfluidity type are
conveniently parametrized by the dependence of the corresponding
critical temperature $T_\mathrm{crit}$ on the mean baryon number density
$n_\mathrm{b}$ or the mass density $\rho$ (see, e.g.,
\citealt{Yakovlev_01}, for review and references). In
Figs.~\ref{fig:qemxsf25gs}--\ref{fig:qemxBSk24BSk25} we used the
parametrizations MSH, BS and TTav by \citet{Ho_15} to the microscopic
calculations by
\citet*{MargueronSH08,BaldoSchulze07,TakatsukaTamagaki04} for the ns, ps
and nt pairing types, respectively. The corresponding density
dependences of $T_\mathrm{crit}$ are shown in Fig.~\ref{fig:tcrit}.

\citet{MargueronSH08} and \citet{Gandolfi_09} reported similar
theoretical dependences of $T_\mathrm{crit}$ on baryon density for the
ns-type pairing. Previously we have checked that the replacement of one
of these models by the other does not appreciably change the thermal
evolution of the isolated neutron stars \citep{PotekhinChabrier18} and
the accreting neutron stars in the SXTs
\citep{PotekhinCC19,PotekhinChabrier21}.  Results of recent numerical
simulations of the neutron singlet superfluidity by \citet{Ding_16} are
also close to the MSH model.

\citet{BaldoSchulze07} showed that the allowance for the three-body
effective nuclear forces and for in-medium effects (such as modified
effective nucleon masses and polarization) reduces the gap in the proton
energy spectrum caused by the ps-pairing, $\Delta_\mathrm{ps}$, which is
related to the critical temperature of superfluidity, $\kB
T_\mathrm{crit} = 0.5669\,\Delta_\mathrm{ps}$. \citet{Guo_19} have
argued that a proper account of the density dependence of the
proton-proton induced interaction in the neutron star matter leads to
still stronger suppression of the ps-type superfluidity. The values of
$T_\mathrm{crit}$ that follows from their tabular values of
$\Delta_\mathrm{ps}$ are shown in Fig.~\ref{fig:tcrit} by the triangles.
\citet{LimHolt21} have recently studied proton pairing in neutron stars
using the chiral effective field theory and obtained a broad variety of gaps
as functions of the proton Fermi momentum $k_\mathrm{F}^\mathrm{p}$,
depending on the details of the theoretical model. All these gaps are
either similar to or smaller than the BS gap.
\citet{SedrakianClark19} suggested that the absence of proton superconductivity may
have profound implications for the physics of compact stars. We tested
the effect of the reduction of the ps-type critical temperature by using
the extreme model of completely suppressed proton superfluidity and
found that it almost does not affect the heating curves shown in
Figs.~\ref{fig:qemxsf25gs}, \ref{fig:qemxsf25He},
\ref{fig:qemxBSk24BSk25}. This is explained by the fact that the
non-suppressed BS proton pairing gap vanishes at the centres of all
neutron stars with $M>\msun$, hence a substantial part of the core is
free of the proton superfluidity in all considered models.

\begin{figure}
\includegraphics[width=\columnwidth]{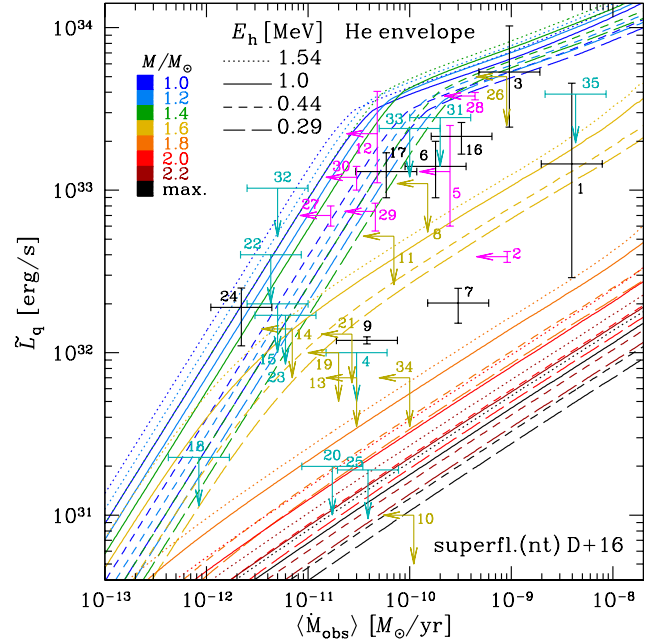}
\caption{The same as in Fig.~\ref{fig:qemxsf25He}, but with another model 
of the neutron triplet-type superfluidity  (D+16 instead
of TTav).
}
\label{fig:qemxsf39He}
\end{figure}

However, the current uncertainties in the theory of the triplet pairing
of baryons prove to be essential for the simulations of the thermal
evolution of neutron stars (see, e.g.,
\citealt{Fortin_18,PotekhinCC19,Potekhin_20,Burgio_21} and references
therein). The main effect of the superfluidity in the core on the
quasi-stationary states and long-term evolution of the neutron stars
consists in suppressing the direct Urca processes. Due to this effect,
the decrease of the luminosity with increasing mass above
$M_\mathrm{DU}$ is less sharp than it would be in the absence of
superfluidity. As an example alternative to the TTav model, the
short-dashed line in Fig.~\ref{fig:tcrit} represents the parametrization
of \citet{Ding_16} to their numerical results, based on the same Av18
effective nucleon-nucleon potential as implied in the TTav model, with
short- and long-range neutron correlations included (hereafter, D+16).
Fig.~\ref{fig:tcrit} reveals that the TTav and D+16 models predict
similar $T_\mathrm{crit}(\rho)$ dependences at relatively small $\rho$
up to the maximum
$T_\mathrm{crit}^\mathrm{(max)}\approx(5-7)\times10^8$~K at
$\rho\approx4\times10^{14}$ \gcc, but at higher densities the D+16
superfluidity becomes quickly suppressed, unlike the TTav one. The D+16
parametrization is used in Fig.~\ref{fig:qemxsf39He}, which should be
compared with Fig.~\ref{fig:qemxsf25He}, where we employed the TTav
model using the same models of the EoS and the heat-blanketing envelope.
Because of vanishing D+16 superfluidity at $\rho\gtrsim7\times10^{14}$
\gcc, any neutron star with $M>M_\mathrm{DU}$ has some central region
where the direct Urca process is not suppressed by superfluidity. This
causes a sharper lowering of the heating curves (i.e., decrease of
$\tilde{L}_\mathrm{q}$ with increasing $M$ at a given
$\langle\dot{M}_\mathrm{obs}\rangle$) in Fig.~\ref{fig:qemxsf39He} relative to
Fig.~\ref{fig:qemxsf25He} for $M > M_\mathrm{DU}$. 

Let us note that the parametrization by \citet{Ding_16} is based on
their computations without inclusion of three-nucleon forces (3NF). The
preliminary results with inclusion of the 3NF effects, presented by
these authors, show an increase of the triplet gap (in contrast with the
singlet pairing case) and do not demonstrate the gap closure at large
densities, so that the TTav gap appears to be between the results of
\citet{Ding_16} with and without the 3NF. The authors caution that at
these high densities one approaches the limit of applicability of the
employed theory.

\begin{figure*}
\includegraphics[width=.5\textwidth]{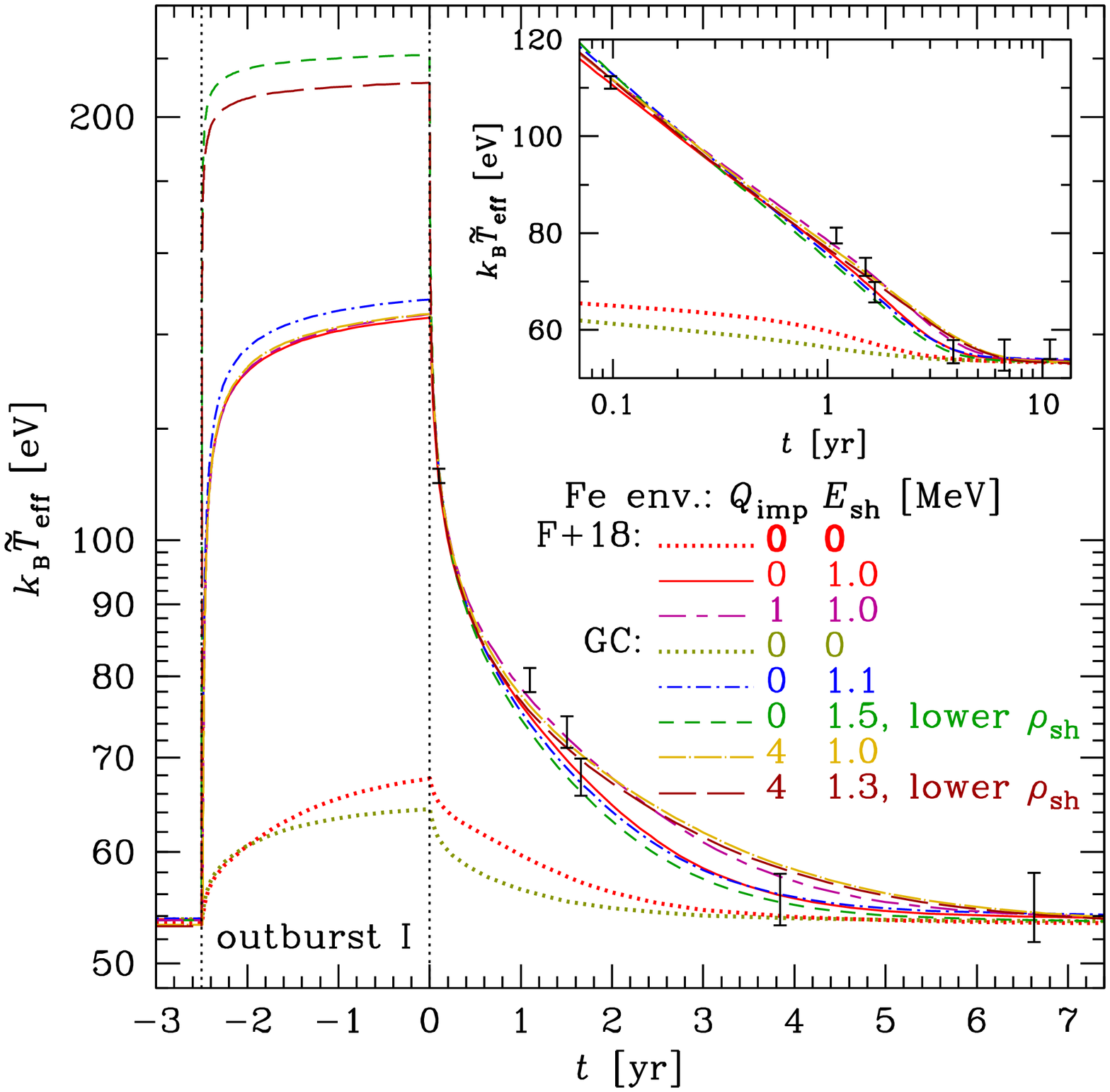}
\includegraphics[width=0.28\textwidth]{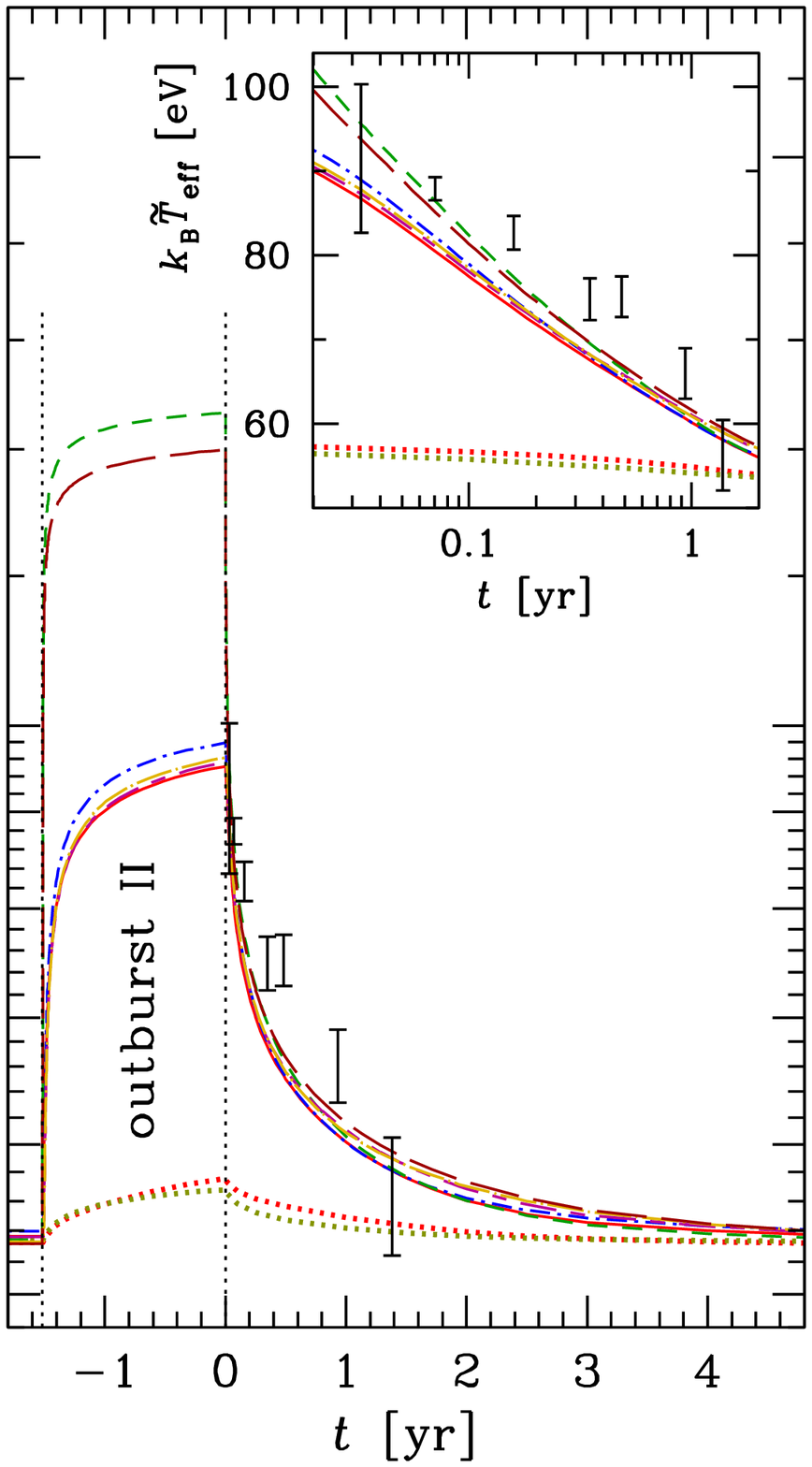}
\caption{Simulated light curves for the outbursts I (left panel) and II 
(right panel) of MXB 1659$-$29 versus observations. The shown 
dependences of the effective temperature in energy units, $\kB
\tilde{T}_\mathrm{eff}$, on time $t$, as measured by a distant observer
from the end of an outburst, have been computed using the BSk24 EoS
model for a neutron star with $M=1.6\,\msun$ with an iron outer
envelope under different assumptions about the accreted crust. The upper
dotted curve, solid curve and long-dash---short-dash curve are
calculated assuming the F+18 model, and the other curves illustrate the
case of the GC model with $P_\mathrm{oi}=P_\mathrm{oi}^\mathrm{(0)}$.
The dotted lines show the case of no shallow heating and the other lines
include the amount of heat per accreted nucleon $E_\mathrm{sh}$ as
quoted in the legend.  The mass density of the shallow heating is taken
to be $\rho_\mathrm{sh}=1.4\times10^9$ \gcc{} for all the curves except
the ones marked `lower $\rho_\mathrm{sh}$' in the legend: for these two
curves $\rho_\mathrm{sh}=10^8$ \gcc. The impurity parameter
$Q_\mathrm{imp}$ is also listed in the legend for each case. The rising
parts of the curves at  $t<0$ show the increase of conditional effective
temperature during an outburst, calculated neglecting the
accretion-induced processes near the surface. The error bars represent
the values derived from observations, according to \citet{Parikh_19}.
The insets show the crust cooling curves plotted at a logarithmic time scale.
}
\label{fig:MXB1659Fe}
\end{figure*}

\section{Crust cooling: the case of MXB 1659--29}
\label{sect:MXB}

\begin{figure*}
\includegraphics[width=.5\textwidth]{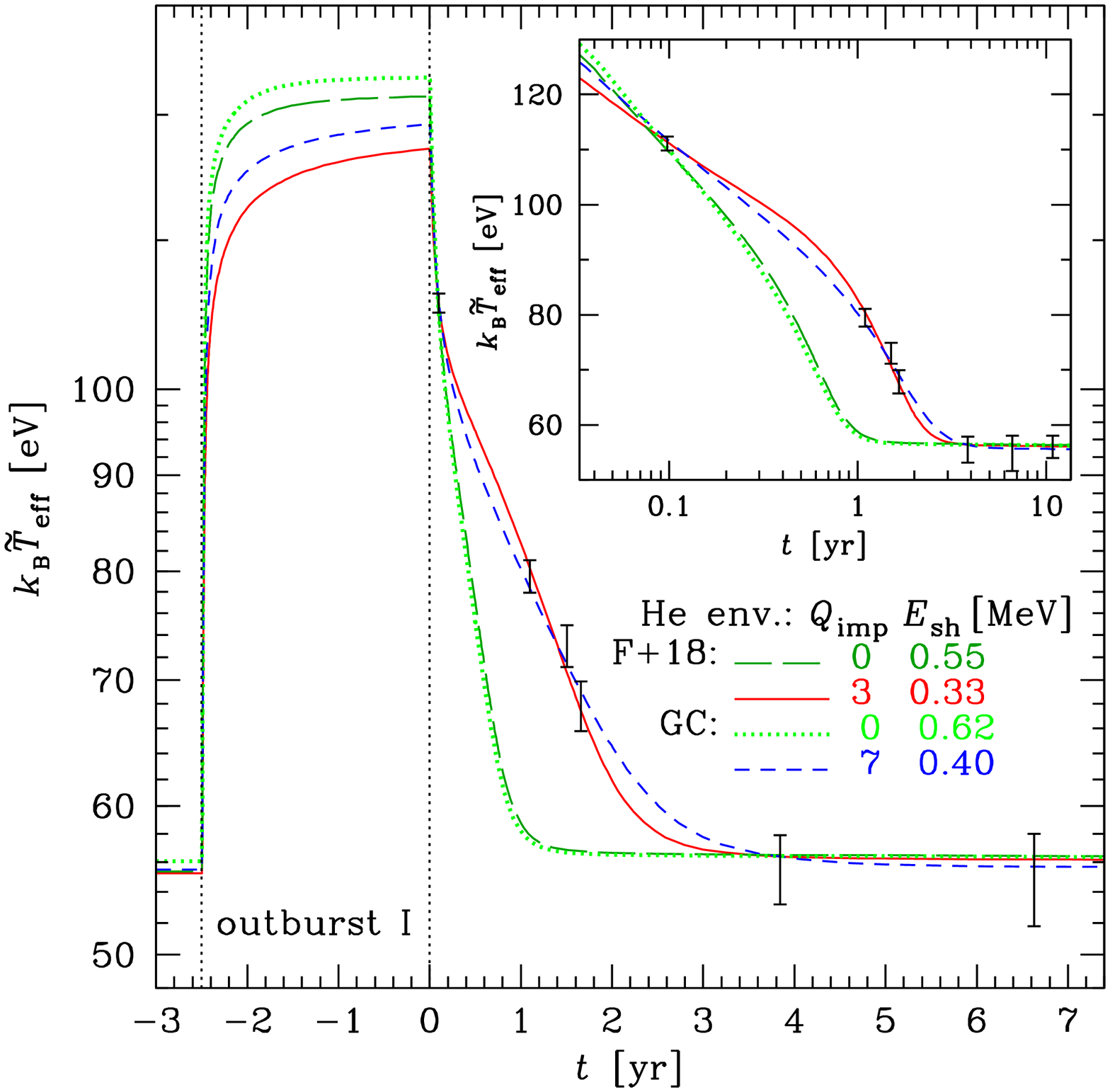}
\includegraphics[width=0.28\textwidth]{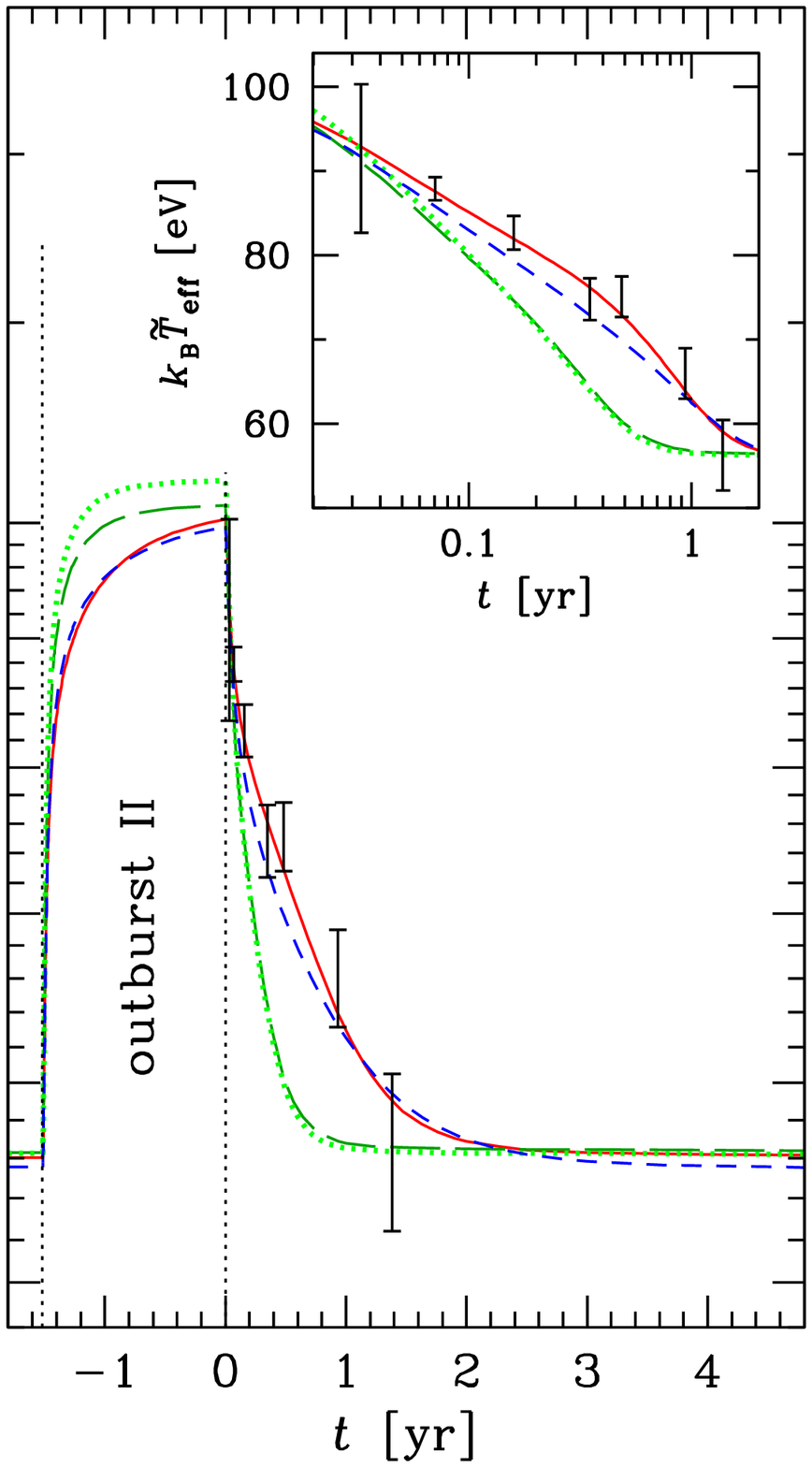}
\caption{The simulated light curves and observational error bars,
analogous to Fig.~\ref{fig:MXB1659Fe}, but for a neutron star of mass
$M=1.8\,\msun$ with the helium atmosphere and ocean down to the
interface with the accreted crust at mass density $\rho=10^8$ \gcc. The
curves of different styles correspond to different $Q_\mathrm{imp}$ and
$E_\mathrm{sh}$ in the crust, according to the legend.
}
\label{fig:MXB1659He}
\end{figure*}

Let us apply the accreted crust models described in
Section~\ref{sect:accrust} to a study of the thermal evolution of the SXT
MXB 1659$-$29 (MAXI J1702$-$301).  We have chosen this quasi-persistent
transient as an example, because it has been observed for a long time,
revealed three outbursts and has a well documented record of the crust
cooling after the last two of them. It was discovered as an X-ray
bursting source during the \textit{SAS-3} satellite mission
\citep{Lewin_76,LewinJoss77}. Later it was observed many times using
different instruments (see \citealt{Wijnands_03,Parikh_19} and
references therein). The source was detected several times in X-rays 
and in the optical from October 1, 1976 till July 2, 1979, but then
(between July 2 and July 17, 1979; \citealt*{Cominsky_OL83}) it turned
to quiescence and could not be detected any more until April 1999, when
\citet{intZand_99} found it to be active again. The source remained
bright for almost 2.5 years before it became dormant in September 2001.
It was first observed in the quiescent state by \textit{Chandra} in
October 2001 \citep{Wijnands_03}. Afterwards, its X-ray emission was
observed several times in quiescence till October 2012
\citep{Cackett_06,Cackett_08,Cackett_13}. In August 2015 the source
showed a new outburst \citep{Negoro_15}, which lasted $\approx550$ days
till February 2017. Subsequent crust cooling was followed from March
2017 using X-ray observatories \textit{Swift}, \textit{Chandra} and
\textit{XMM-Newton}. The results have been summarized and analyzed by
\citet{Parikh_19}. Following these authors, we name \emph{outburst I}
and \emph{outburst II} those of 1999--2001 (MJD 51265--52162) and
2015--2017 (MJD 57256--57809.7), respectively; we also name
\emph{outburst 0} the one observed in 1976--1979. The quiescent light
curves after the end of outburst I were modeled in a number of works
\citep{BrownCumming09,Cackett_13,Deibel_17,Parikh_19,PotekhinChabrier21,Mendes_22}.
\citet{Parikh_19} and \citet{PotekhinChabrier21} performed consistent
modelling of the short-term evolution of MXB 1659$-$29 during and after
the two  outbursts I and II. In all the cases, additional adjustable
model parameters were necessary for good fits: the heat deposited
at shallow depths per accreted baryon, $E_\mathrm{sh}$, and the
mean-square deviation of the nuclear charge number due to impurities in
the crust,  $Q_\mathrm{imp}$. The latter parameter modifies the electron
thermal conductivity of the crust: the larger $Q_\mathrm{imp}$, the
smaller the conductivity.

The previous studies were based on the traditional models of the nuclear
transformations and heat production in the accreted crust. Here we use
the case of MXB 1659$-$29 to illustrate the influence of 
neutron
redistribution
in the inner crust on the post-outburst cooling of neutron
stars in SXTs. For this purpose we will compare the results of 
numerical simulations of the thermal evolution of the neutron star in
this SXT, performed using the traditional (F+18) and thermodynamically
consistent (GC) accreted crust models. For each model, we first prepare
a quasi-equilibrium thermal state of a neutron star so as to match the
most probable value of the redshifted effective temperature
$\tilde{T}_\mathrm{eff}$ in quiescence, derived from observations,
$\kB\tilde{T}_\mathrm{eff}\approx 56$ eV
\citep{Cackett_08,Cackett_13}.\footnote{Following the previous works
\citep{Parikh_19,PotekhinChabrier21,Mendes_22}, here we discard the
alternative estimate $\kB\tilde{T}_\mathrm{eff}\approx 43$ eV obtained
by  \citet{Cackett_13} with inclusion of a power-law spectral component
in addition to the thermal one.} Then we consecutively simulate the
neutron star heating during outburst 0, followed by cooling during 20
years, heating during outburst I, followed by cooling during 13.9 years
and heating during outburst II, followed by cooling till now. In the
absence of detailed information on the early outburst 0, we assume that
it has the same characteristics as outburst~I. We fix
$\dot{M}=4\times10^{-9}\,\msun$ yr$^{-1}$ for the outburst I and
$\dot{M}=1.3\times10^{-9}\,\msun$ yr$^{-1}$ for outburst~II
\citep[see][for details]{PotekhinChabrier21}.

First let us consider the heavy-element (Fe) heat-blanketing envelope
model. In this case, as can be seen from Figs.~\ref{fig:qemxsf25gs}
and~\ref{fig:qemxAPR}, it can be compatible with the estimated
$\langle\dot{M}\rangle$ and $\tilde{L}_\mathrm{q}$ only if $M\approx
M_\mathrm{DU}$. For illustration we assume the BSk24 EoS and accordingly
fix $M=1.6\,\msun$. The results are shown in Fig.~\ref{fig:MXB1659Fe}.
The upper and lower dotted lines correspond to the F+18 and GC models,
respectively, without any adjustable ingredients, i.e., with
$E_\mathrm{sh}=0$ and $Q_\mathrm{imp}=0$. We see that they cannot
reproduce the amplitude of the observed crust cooling. The other light
curves in this figure are calculated by adjusting $E_\mathrm{sh}$ so as
to match the first measurement of $\tilde{T}_\mathrm{eff}$ after the end
of outburst~I (the \textit{Chandra} observation on October 15, 2001,
about one month after the source became quiescent). 
We find that
we need to
include $E_\mathrm{sh}\approx 1$ MeV, if it is located near the
shallowest nuclear transformation in the F+18 and GC models of accreted
crust, $\rho_\mathrm{sh}\approx1.4\times10^9$ \gcc{} 
(that is near the inner
boundary of the region where burning of light elements to iron can be
compatible with the data in Tables \ref{tab:GCheat24} and
\ref{tab:GCheat25}), which is our default value. If the shallow heating
occurs at a smaller density, then the heat amount $E_\mathrm{sh}$ should
be somewhat larger, because its larger part leaks to the surface and
does not heat the inner layers (cf.{} Fig.~\ref{fig:fluxfrac}). For
example, the dashed curves in Fig.~\ref{fig:MXB1659Fe} illustrate the
case where  $\rho_\mathrm{sh}=10^8$ \gcc. A comparison with the case of
the default shallow depth (the dot-dashed curve) shows that the crust
cooling is rather insensitive to the shallow heating depth at
$t\gtrsim0.1$ yr, but the smaller $\rho_\mathrm{sh}$ gives a better
agreement with the observations of outburst~II at $t\lesssim 0.1$ yr
(see the inset in the right panel of Fig.~\ref{fig:MXB1659Fe}).

An admixture of charge impurities helps us to achieve a better agreement
between the theoretical crust cooling curves and observations. For
simplicity, we assume constant impurity level $Q_\mathrm{imp}$ in the
entire solid crust. The optimal values are  $Q_\mathrm{imp}\approx 1$
and  $Q_\mathrm{imp}\approx 4$ in the F+18 and GC models, respectively.

\begin{figure}
\includegraphics[width=\columnwidth]{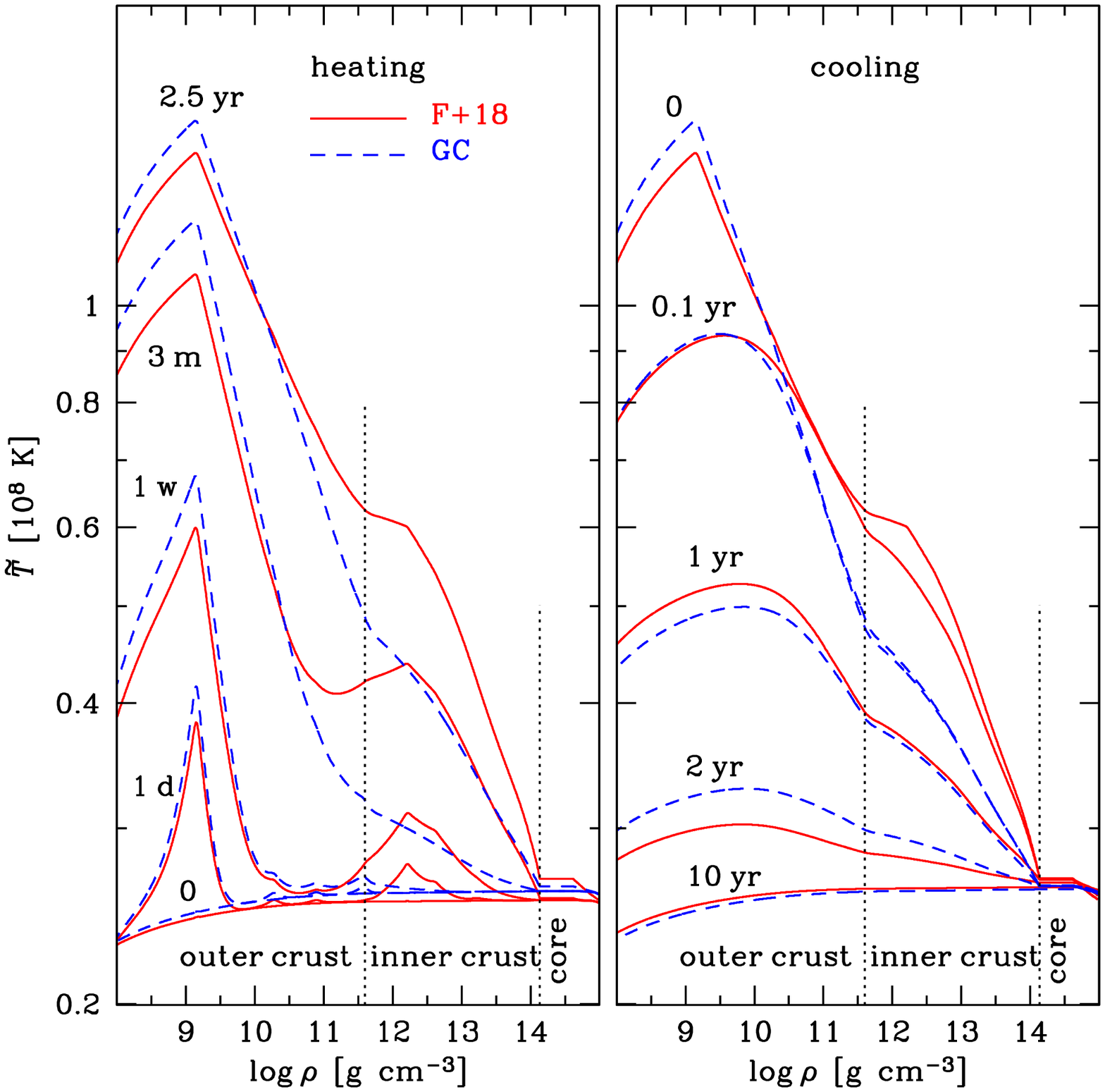}
\caption{Redshifted temperature profiles inside a neutron star as in
Fig.~\ref{fig:MXB1659He} during the outburst~I (left panel) and
subsequent cooling (right panel) at different times after the start of
the heating and the cooling, respectively, as marked near the curves.
The same neutron star model as in Fig.~\ref{fig:MXB1659He} is used. The
solid and dashed curves  represent the results for the accreted crust
models F+18 and GC with the same accretion rate, shallow heating and
impurity parameters as for the respective curves in the left panel of
Fig.~\ref{fig:MXB1659He}. The vertical dotted lines indicate the
outer-inner catalyzed crust and crust-core interfaces.
}
\label{fig:profiles}
\end{figure}

\begin{figure}
\includegraphics[width=.5\textwidth]{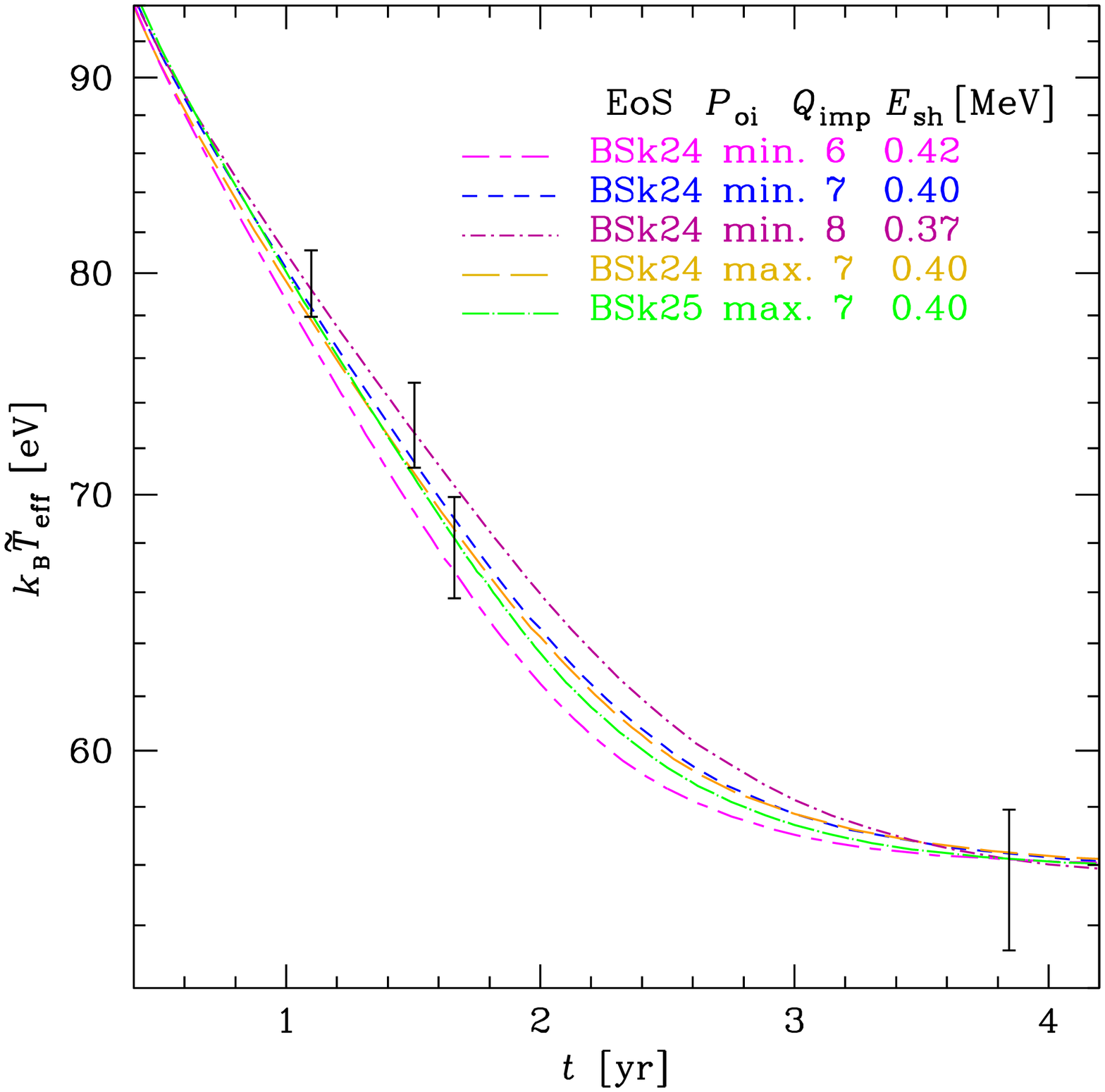}
\caption{Comparison of the best-fit simulated light curve for the GC
model from Fig.~\ref{fig:MXB1659He}, left panel (the blue short-dashed
line) with light curves computed, as clarified in the legend, for
different $Q_\mathrm{imp}$ and $E_\mathrm{sh}$ in the crust and for the
model with $P_\mathrm{oi}=P_\mathrm{oi}^\mathrm{(cat)}$ (marked `max.'
in the legend, whereas the choice
$P_\mathrm{oi}=P_\mathrm{oi}^\mathrm{(0)}$ is marked `min.'), for the
models BSk24 and BSk25.
}
\label{fig:MXB1659var}
\end{figure}

Let us consider another model of the heat-blanketing envelope, which is
composed of He at $\rho<10^8$ \gcc{} and is consistent with the adopted
accreted crust model (either GC or F+18) at $\rho>10^8$ \gcc. This
envelope is slightly less heat-transparent than the He envelope
considered in Section~\ref{sect:long}. Keeping the BSk24 EoS in the
core, we find that the realistic mean accretion rates $\langle
\dot{M}_\mathrm{obs}\rangle \sim 10^{-10}-10^{-9}\,\msun$ yr$^{-1}$ are
compatible with the estimated surface temperature in quiescence,
$\tilde{T}_\mathrm{eff}=56\pm2$ eV \citep{Cackett_13,Parikh_19}, if
$M\approx(1.7-2.0)\,\msun$, depending on the nt-superfluidity model
($\approx2.0\,\msun$ for the TTav model and $\approx1.7\,\msun$ for the D+16
model; cf.\ Figs.~\ref{fig:qemxsf25He} and \ref{fig:qemxsf39He}). The
results of the simulations with fixed $M=1.8\,\msun$ are shown in
Fig.~\ref{fig:MXB1659He}. Here, all light curves are calculated with
$E_\mathrm{sh}$ adjusted so as to match the first measurement of
$\tilde{T}_\mathrm{eff}$ after the end of outburst~I. Assuming
$Q_\mathrm{imp}=0$, we obtain very similar light curves for the F+18 and
GC models,which exhibit a fast post-outburst cooling incompatible with
the observations.  We see that an acceptable agreement between the
theory and observations can be achieved if we assume
$Q_\mathrm{imp}\approx3$ in the F+18 model or $Q_\mathrm{imp}\approx7$
in the GC model. It is remarkable that the observations of both
outbursts I and II can be fitted by using the same values of
$E_\mathrm{sh}$ and $Q_\mathrm{imp}$.\footnote{We recall that the
consistent fitting of the cooling light curves after both outbursts I
and II was previously performed by \citet{Parikh_19} and
\citet{PotekhinChabrier21} using the traditional accreted crust models.}

The presented results demonstrate that the F+18 model provides somewhat
better agreement with the observations than the GC model, although this
difference is statistically insignificant. It is due to the retarded
cooling in the F+18 model at intermediate times $t \sim (0.1-1)$ yr,
which is absent in the GC model. Fig.~\ref{fig:profiles} provides an
insight into the origin of this retardation. In the traditional model,
unlike the GC model, the inner crust layers at $\rho\sim10^{12}-10^{13}$
\gcc{} are heated appreciably during the outburst. This inner-crust heat
does not affect the initial cooling stage, as we see from the near
constancy of $\tilde{T}$ at densities of a few times $10^{11}$ \gcc{} at
$t \lesssim 0.1$ yr. At $t\gtrsim0.1$ yr, however, this additional heat
propagates to the surface and is radiated, making the luminosity
somewhat higher than in the GC model. In the right panel of
Fig.~\ref{fig:profiles}, the temperature profiles at $\rho\lesssim10^9$
\gcc{} are lower in the F+18 model than in the GC model at early times
$t<0.1$ yr because of the stronger preceding shallow heating, but become
higher at $t\sim1$ yr due to the heat income from the inner
crust. At $t\gtrsim2$ yr the GC profiles become higher again because the
heat dissipates less quickly due to the larger impurity parameter (i.e.,
lower conductivity of the crust).

Fig.~\ref{fig:MXB1659var} illustrates theoretical variations of the SXT
crust cooling behaviour in frames of the GC models. First, we vary
$Q_\mathrm{imp}$ around the best-fit GC model shown in
Fig.~\ref{fig:MXB1659He} and adjust $E_\mathrm{sh}$ as previously. We
see that the observational error bars allow variation of
$Q_\mathrm{imp}$ by $\sim20$\%, an increase of $Q_\mathrm{imp}$ being
accompanied by a decrease of $E_\mathrm{sh}$. Next, we test the model
with the largest $P_\mathrm{oi}$ value
($P_\mathrm{oi}=P_\mathrm{oi}^\mathrm{(cat)}$), instead of the smallest
one ($P_\mathrm{oi}=P_\mathrm{oi}^\mathrm{(0)}$). Finally, we replace the
BSk24 model for the heating profile and crust composition
(Table~\ref{tab:GCheat24}) by the BSk25 model
(Table~\ref{tab:GCheat25}).
 As we see, neither the variation of
$P_\mathrm{oi}$ nor the replacement of BSk24 by BSk25 can affect the
crust cooling appreciably. This is explained by the
dominant influence of the heat sources in the outer crust (including
the shallow heating), which are the same in these models.

\section{Conclusions}
\label{sect:concl}

We studied
the consequences of
hydrostatic/diffusion equilibrium
of free neutrons in accreted
inner crust on the long- and short-term  thermal
evolution of the neutron stars in 
the 
SXTs. We found that the observed
quasi-stationary thermal luminosities of such neutron stars can be
equally well fitted in the frames of the model F+18 \citep{Fantina_18},
based on the traditional approach that neglects 
the hydrostatic/diffusion equilibrium condition for free neutrons
and the models GC \citep{GusakovChugunov20,GusakovChugunov21}, which
take this condition into account.
However, in the latter model, unlike the traditional ones, good fits of
some relatively hot quiescent sources can only be obtained provided that
the heating of the neutron star core by the energy sources located at
shallow depths in the outer crust is allowed for.
We also
found that the observations of the cooling of SXT MXB 1659$-$29 after
its two consecutive outbursts can be consistently fitted in both the
F+18 and GC models, but using the adjustable parameters of the shallow
heating and crust impurities that are different for these models.

\section*{Acknowledgments}

We are grateful to Nikolai Shchechilin for a useful remark and to the
anonymous referee for valuable comments.  MEG is grateful to the
Department of Particle Physics \& Astrophysics at the Weizmann Institute
of Science for hospitality.
The work was supported by the Russian Science Foundation grant
22-12-00048.

\section*{Data availability}
The data underlying this article will be shared on
reasonable request to the corresponding author.

\newcommand{\arnps}{Annu.\ Rev.\ Nucl.\ Part.\ Sci.}
\newcommand{\al}{Astron.\ Lett.}
\newcommand{\jaa}{JA\&A}
\newcommand{\ptp}{Prog.\ Theor.\ Phys.}
\newcommand{\ppnp}{Prog.\ Part.\ Nucl.\ Phys.}
\newcommand{\rmp}{Rev.\ Mod.\ Phys.}

\label{lastpage}

\end{document}